\documentclass[doublecol]{epl2}

\usepackage{epsfig}
\usepackage{graphicx}
\usepackage{dcolumn}
\usepackage{bm}

\usepackage{ulem}

\usepackage[dvipdfm]{hyperref}

\title{Electronic structure of La$_2$CoSi$_3$ - a non-Kondo analogue of
a Kondo lattice, Ce$_2$CoSi$_3$}
\author{Swapnil Patil, Garima Saraswat, Ganesh Adhikary, R. Bindu,
E. V. Sampathkumaran and Kalobaran Maiti \footnote{Corresponding
author: kbmaiti@tifr.res.in}}

\institute{Department of Condensed Matter Physics and Materials'
Science, Tata Institute of Fundamental Research, Homi Bhabha Road,
Colaba, Mumbai 400005, India}
\date{\today}
\abstract {We study the electronic structure of a Pauli paramagnetic
compound, La$_2$CoSi$_3$ using photoemission spectroscopy and {\it
ab initio} band structure calculations. Experimental valence band
spectra exhibit signature of electron correlation induced feature
around 2.5 eV - the correlation strength among Co 3$d$ electrons is
estimated to be close to 3 eV. The Co 2$p$ core level spectra also
exhibit correlation induced satellite features consistent with the
scenario in the valence band spectra suggesting importance of
conduction electron correlation in addition to the local moment in
Kondo lattice systems. The La$_2$CoSi$_3$ valence band spectra could
be utilized to extract Ce 4$f$ related spectral features and thus
provide a good reference to study Kondo lattice systems in this
class of materials. Temperature evolution of various core level
spectra is found to be complex revealing deviations from a typical
Fermi liquid behavior and emergence of distinct surface-bulk
difference in the electronic structure at finite temperature.}

\begin{document}

\maketitle

\section{Introduction}

Kondo physics has been one of the most studied subjects in the
contemporary condensed matter physics for many
decades.\cite{Brandt,thomas} A lot of emphasis is given on Ce-based
intermetallic compounds for their varied electronic properties such
as valence fluctuations, Kondo screening, heavy fermion
superconductivity etc. Such properties are known to arise due to the
subtle tuning of the hybridization between the Ce 4$f$ electrons and
the valence electrons. The theoretical description of such local
moment systems is usually captured employing Anderson impurity
models.\cite{thomas,Anderson} An important input in such
descriptions is the correct estimation of the valence band spectral
functions for the calculation of the hybridization induced
effects.\cite{Wills,Gunnarsson1,Gunnarsson2,Gunnarsson3} There have
been efforts to utilize various model spectral functions as an
approximation for such a description. In many cases, photon energy
dependence of the photoemission cross section of various electronic
states has been utilized to experimentally extract Ce 4$f$ related
partial density of states.\cite{arko}

\begin{figure}
\includegraphics [scale=0.4]{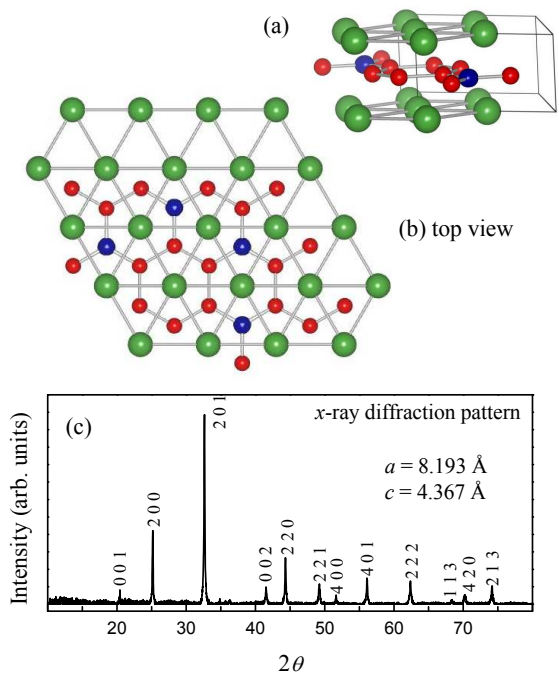}
\vspace{-16ex}
 \caption{(a) Crystal structure of La$_2$CoSi$_3$. (b) Top view of the same
crustal structure to visualize the atomic arrangements in each layer
with clarity. (c) $X$-ray diffraction pattern of La$_2$CoSi$_3$.}
 \vspace{-2ex}
\end{figure}

The electronic structure of a unoccupied 4$f$ homologue of the
Ce-based systems can be exploited to extract the parameters more
precisely pertinent to the Ce 4$f$-valence states hybridization in a
realistic scenario. While any of the Ce-based system can be
considered for such investigations, we chose here '213'
($R_2$CoSi$_3$; $R$ = rare-earths) class of materials,\cite{Gordon}
which can be formed easily in single phase, one can continuously
tune the electronic properties from Pauli paramagnetic phase to
Kondo lattice limit via a single impurity regime. La$_2$CoSi$_3$
forms in AlB$_2$ type hexagonal structure as shown in Fig. 1(a) and
1(b), where La forms hexagonal layers sandwiched by (CoSi$_3$)
layers - top view shown in Fig. 1(b) depicts the atomic arrangements
in the layers. Among large number of materials studied in this
class, Ce$_2$CoSi$_3$ is well established to be a concentrated Kondo
lattice system.\cite{sampath1} La substitution in Ce$_2$CoSi$_3$
helps to dilute the Ce-density thereby pushing the system towards
single impurity regime. It has been observed that strong Kondo
coupling strength in this systems necessitates a large degree of
substitution ($>$ 50\% of Ce density) to achieve deviation from the
typical Kondo lattice behavior.\cite{sampath2} The end member
La$_2$CoSi$_3$ is a Pauli paramagnetic metal and therefore, an ideal
compound to provide a testing ground for the studies of Kondo
lattice system in this interesting class of materials.

We employed $x$-ray photoemission and ultraviolet photoemission
spectroscopy to directly probe the electronic structure and discuss
the results in comparison with those for Ce$_2$CoSi$_3$. Our results
reveal interesting electronic structure with significant anomaly
albeit this compound being Pauli paramagnetic. The comparative study
helps to identify effects due to the electron correlation involving
Ce 4$f$ states distinct from the Fermi liquid behavior.
Additionally, the core level spectra show unusual evolution with
temperature.

\section{Experimental details}

La$_2$CoSi$_3$ was prepared by melting together stoichiometric
amounts of high purity ($>$ 99.9\%) La, Co and Si in an arc furnace
in the atmosphere of high purity argon. The sample was annealed at
750 $^o$C for one week and then characterized by $x$-ray diffraction
(XRD). The XRD pattern, shown in Fig. 1(c), exhibits sharp
diffraction peaks corresponding to the AlB$_2$-derived hexagonal
structure (space group $P6/mmm$)\cite{Gordon} suggesting the sample
to be in single phase.

The photoemission measurements were performed using a Gammadata
Scienta R4000 analyzer and monochromatic laboratory photon sources
at an energy resolutions set to 0.4 eV and 2 meV at Al $K\alpha$
(1486.6 eV) and He {\scriptsize II}$\alpha$ (40.8 eV) photon
energies, respectively. The base pressure in the vacuum chamber was
3 $\times$ 10$^{-11}$ Torr. The temperature variation down to 10~K
was achieved by an open cycle He cryostat (LT-3M, Advanced Research
Systems, USA). The clean surface of the melt grown ingots (extremely
hard to cleave) was obtained by {\it in situ} fracturing and/or
scraping - the surface cleanliness was ensured by the absence of
O~1$s$ and C~1$s$ features in the $x$-ray photoelectron spectra and
the absence of impurity features in the binding energy range of 5-6
eV in the ultraviolet photoelectron spectra. Both the surface
preparation procedures resulted similar data. The reproducibility of
the spectra was confirmed after each trial of cleaning process.


The electronic band structures corresponding to the experimentally
found ground state were calculated employing {\it state-of-the-art}
full potential linearized augmented plane wave (FLAPW) method using
WIEN2k software\cite{wien} within the local density approximations,
LDA. The convergence for different calculations were achieved
considering 512 $k$ points within the first Brillouin zone. The
lattice constants used in these calculations are determined from the
$x$-ray diffraction patterns ($a$~=~8.193~\AA\ and $c$ =~4.367~\AA)
for La$_2$CoSi$_3$. The lattice constants for Ce$_2$CoSi$_3$ were
taken from elsewhere.\cite{Gordon}

\section{Results and discussions}

\begin{figure}
 \vspace{-2ex}
\includegraphics [scale=0.4]{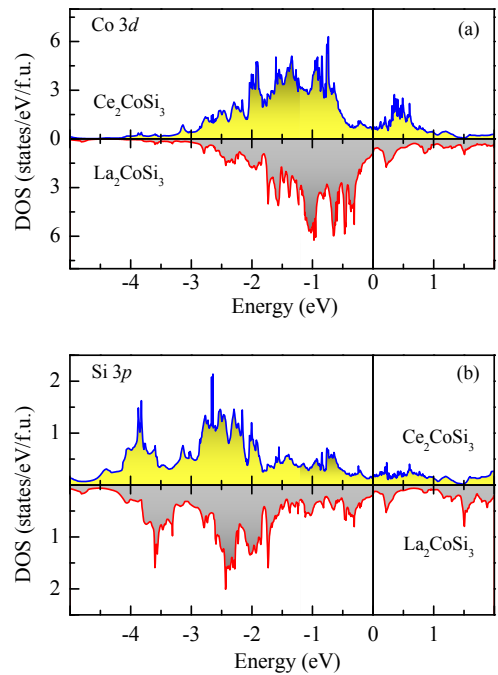}
\vspace{-10ex}
 \caption{Calculated (a) Co 3$d$ and (b) Si 3$p$ partial density of states
of Ce$_2$CoSi$_3$ and La$_2$CoSi$_3$ within local density
approximation (LDA).}
 \vspace{-2ex}
\end{figure}

The calculated Co 3$d$ and Si 3$p$ partial density of states (PDOS)
for Ce$_2$CoSi$_3$ and La$_2$CoSi$_3$ are shown in Fig. 2(a) and
2(b), respectively. The 4$f$ PDOS are not shown here as these
contributions in La$_2$CoSi$_3$ appear in the unoccupied part of the
electronic structure. The 4$f$ bands in Ce$_2$CoSi$_3$ are barely
occupied. Therefore, the influences of the 4$f$ electrons and the
correlations among them are insignificant in the occupied part of
the electronic structure in the energy scale shown here. The energy
distribution of Co 3$d$ and Si 3$p$ PDOS for both Ce$_2$CoSi$_3$ and
La$_2$CoSi$_3$ resemble well to each other, except there is a rigid
energy shift of about 0.35 eV towards lower energies in
Ce$_2$CoSi$_3$ - this could be attributed to the Fermi level shift
arising from larger electron count of Ce. There is a strong mixing
of the Co 3$d$ and Si 3$p$ states. The density of states near the
chemical potential are the anti-bonding bands having dominant Co
3$d$ character. The bonding bands appear beyond -1.5 eV and possess
large Si 3$p$ character.

\begin{figure}
 \vspace{-2ex}
\includegraphics [scale=0.4]{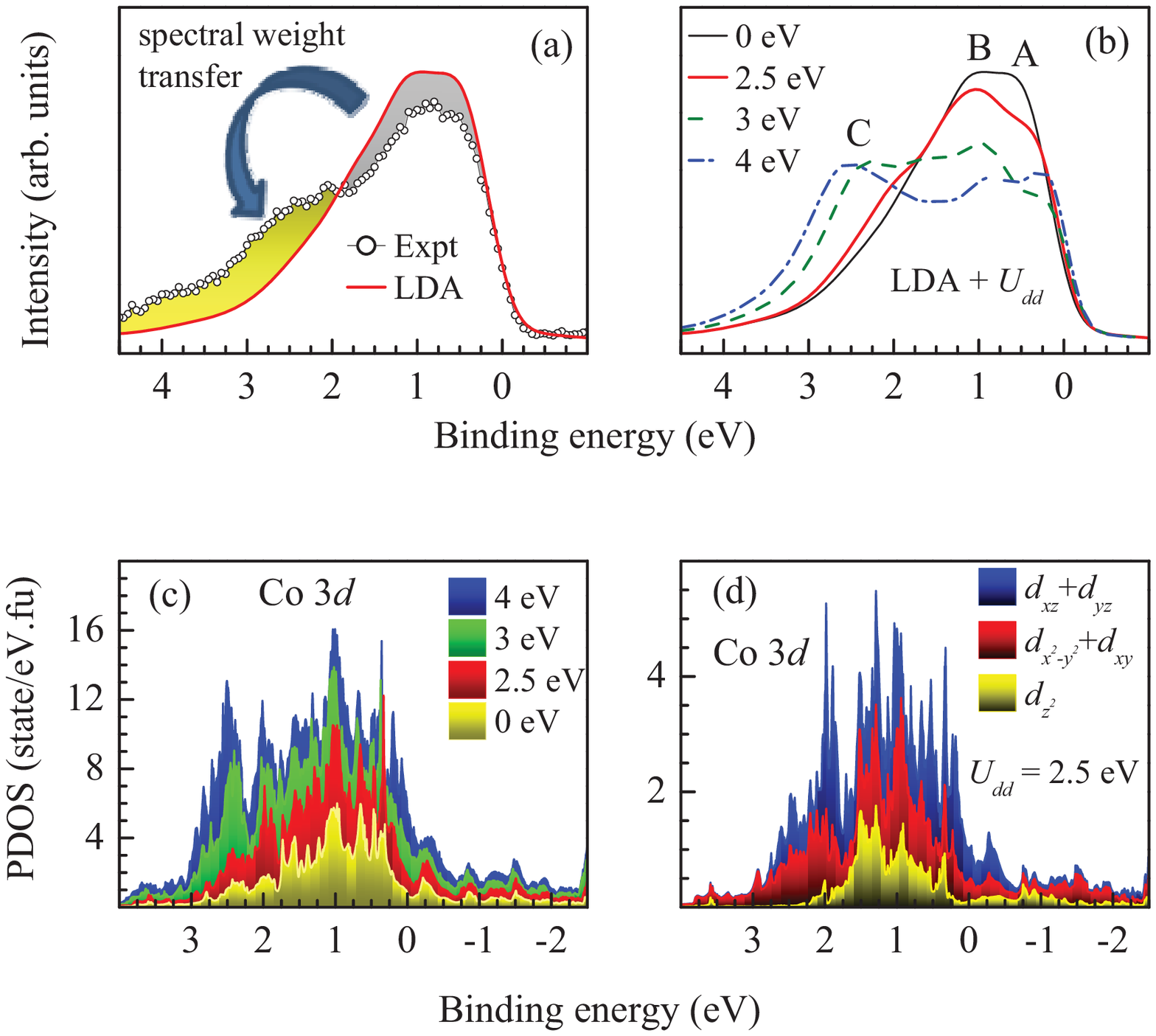}
\vspace{-18ex}
 \caption{(a) The experimental $x$-ray photoemission
spectrum (circles) of the valence band spectrum (circles) of
La$_2$CoSi$_3$ and the spectral function calculated from LDA results
(line). (b) Photoemission spectral function for different values of
$U_{dd}$ obtained from LDA+ $U_{dd}$ results. (c) Stacked area plot
of Co 3$d$ PDOS for various values of $U_{dd}$. (d) Stacked area
plot of Co $3d_{z^2}$, ($3d_{x^2-y^2}+3d_{xy}$) and
$(3d_{xz}+3d_{yz})$ PDOS.}
 \vspace{-2ex}
\end{figure}

In Fig. 3(a), we compare the calculated results with the
experimental valence band spectrum. The integral background
subtracted experimental $x$-ray photoemission (XP) valence band
spectrum for La$_2$CoSi$_3$ is shown by open circles exhibiting
primarily two features centered around 0.5 eV and 2.5 eV binding
energies. The theoretical representation of the valence band spectra
was simulated as follows. The PDOSs for La 5$d$, Co 3$d$ and Si 3$p$
per formula unit were multiplied by the corresponding photoemission
cross-sections for $x$-ray photoemission spectroscopy using 1486.6
eV photon energy.\cite{Yeh} The results were then convoluted by a
Lorentzian function to account for the lifetime broadening and a
Gaussian function to account for the instrumental resolution
broadening. The calculated spectra are shown by lines in the figure.
Evidently, the feature around 0.5 eV is overestimated in the
spectral function corresponding to the LDA results and the feature
around 2.5 eV marked by '$\star$' in the figure is highly
underestimated. Such large difference between the experimental
spectrum and the LDA results can be attributed to the correlation
induced effect among the Co 3$d$ electrons, which are known to be
correlated leading to significant local moment and it is often
underestimated in the LDA calculations. Thus, the experimental
features at 0.5 eV and 2.5 eV can be termed as coherent and
incoherent features, respectively.\cite{lcvo}

In order to capture this scenario of the correlation induced effect
in the electronic structure, we have performed LDA + $U$
calculations for different correlation strength among Co 3$d$
electrons, $U_{dd}$. The simulated valence band spectra for
different values of $U_{dd}$ is shown in Fig. 3(b). The shape of the
coherent feature is better represented in the LDA + $U$ data with
the two distinctly identifiable features A and B having comparable
intensities with the experimental spectrum. An intense feature, C
gradually develops with the increase in $U_{dd}$ value. This is
evident in the Co 3$d$ PDOS shown in Fig. 3(c). Co 3$d$ orbital
contributions for $U_{dd}$ = 2.5 eV are shown in Fig. 3(d)
exhibiting weak influence of $U_{dd}$ on the electrons possessing
3$d_{z^2}$ character while all others exhibit large spectral weight
transfer. The curves corresponding to 2.5 eV and 3 eV in Fig. 3(b)
appear to be very similar to the experimental results providing an
estimate of the Hubbard, $U$ of about 3 eV. This value is quite
similar to the estimations in analogous systems.\cite{patil} Further
increase in $U_{dd}$ leads to much higher binding energy of the
correlation induced feature and a large depletion in intensity
around 1 eV.

It is to note here that all the calculated results presented above
correspond to the experimentally found ground state of both the
materials. It may happen that the approximations in these density
functional approaches may lead to a significantly different magnetic
ground state as found in a classic case of Pu.\cite{Pu} In such a
case, one needs to consider different calculational method such as
LDA+DMFT (DMFT = dynamical mean field theory) to capture the
experimental scenario.\cite{correlation} We hope that the results in
this study would help to initiate further studies in these
directions. Despite simplicity of the method of calculating the
electronic structure followed here, the representation of the
experimental spectra shown in Fig. 3 is remarkable and suggests a
good starting point for the study of these complex systems. Clearly,
the heavy Fermion systems with transition metal $d$ electrons as
conduction electrons are more complex due to the finite correlation
among the conduction electrons.

\begin{figure}
 \vspace{-2ex}
\includegraphics [scale=0.4]{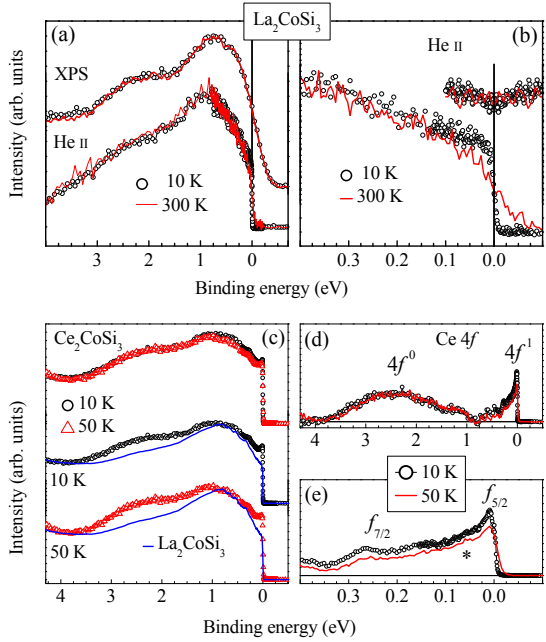}
\vspace{-12ex}
 \caption{(a) Valence band spectra of La$_2$CoSi$_3$ at 10 K and 300 K
obtained by Al $K\alpha$ and He {\scriptsize II} excitations. (b)
The near Fermi level part of the He {\scriptsize II} spectra
obtained employing high energy resolution is shown here for 10 K and
300 K. The spectral density of states obtained by symmetrization are
also shown. (c) He {\scriptsize II} spectra of Ce$_2$CoSi$_3$ at 10
K (circles) and 50 K (triangles) and the corresponding
La$_2$CoSi$_3$ spectra (lines). (d) Extracted Ce 4$f$ spectral
function of Ce$_2$CoSi$_3$ at 10 K and 50 K. (e) Expanded near Fermi
level part of Ce 4$f$ spectra.}
 \vspace{-2ex}
\end{figure}

The above results exemplify the importance of non-$f$ homologue in
the study of heavy Fermion physics in 4$f$ based compounds. We now
investigate the temperature evolution of the valence band spectra of
this non-Kondo system in Fig. 4 - the spectra at 10 K and 300 K
obtained by $x$-ray and He {\scriptsize II} excitations are compared
in Fig. 4(a). The two feature structure observed in the XP spectra
are also found in the He {\scriptsize II} spectra at similar energy
positions. Relatively sharper lineshape of the near Fermi level
feature in the He {\scriptsize II} spectra appears due to the high
energy resolution of the technique. Dominant lifetime broadening at
higher binding energies has smeared out the high resolution induced
effects in the incoherent feature. The spectra away from the Fermi
level exhibit similar lineshape at both the temperatures, 10 K and
300 K.

The features close to the Fermi level in the high resolution He
{\scriptsize II} spectra are shown in Fig. 4(b). The spectra at 10 K
and 300 K exhibit normal Fermi liquid type evolution. To verify this
behavior further, the near $\epsilon_F$ spectra are symmetrized
about $\epsilon_F$ - such a symmetrized spectral intensity often
provides a good representation of the spectral density of states
(SDOS) close to $\epsilon_F$.\cite{Medicherla} The SDOS for
La$_2$CoSi$_3$ does not show any change with temperature
establishing simple Fermi liquid nature of this material.

These spectral functions can be utilized to extract the Ce 4$f$
contributions from the spectra of a Kondo lattice, Ce$_2$CoSi$_3$
having Kondo temperature of about 50 K.\cite{patil1} In Fig. 4(c),
we show the raw data at different temperatures with suitable
normalizations to demonstrate the procedure of Ce 4$f$ extraction.
The high resolution spectra of La$_2$CoSi$_3$ are normalized in such
a way that the subtraction of La$_2$CoSi$_3$ signal from
Ce$_2$CoSi$_3$ spectra does not generate unphysical negative
intensities. Interestingly, this normalization procedure led to
similar background intensities beyond 4 eV. The raw data from
Ce$_2$CoSi$_3$ at 10 K and 50 K are compared in the upper panel of
Fig 4(c) exhibiting temperature dependence as expected due to Kondo
type behavior. The subtracted spectra ($I(Ce_2CoSi_3) -
I(La_2CoSi_3)$), shown in Fig. 4(d) exhibit two peak structure,
typical of the Ce 4$f$ spectral function from hybridized Ce systems,
with one peak close to 2.3 eV binding energy and another peak close
to $\epsilon_F$.\cite{Gunnarsson4,Allen} The feature around 2.3 eV
is denoted by 4$f^0$ and corresponds to the unscreened 4$f$
photoemission ($|4f^1> \rightarrow |4f^0>$ transition). The feature
at the Fermi level, termed $4f^1$ signal appears due to $|4f^1>
\rightarrow |4f^1\underbar{c}>$ transition, where a conduction
electron coupled to the 4$f$ states screens the
photohole.\cite{patil,Ehm} This coupled state contains the
contribution from the Kondo states as evident from the enhancement
of intensity with the decrease in temperature.

The near $\epsilon_F$ part is shown in an expanded scale in Fig.
4(e). There are two distinct feature in the spectra. The feature
close to the Fermi level is denoted by $f_{5/2}$ that corresponds to
the total angular momentum of (5/2) for the charge transferred 4$f$
electron. The feature around 280 meV corresponds to the final state
angular momentum of (7/2). A small kink around 50 meV is also
observed as shown by '$\star$' corresponding to the crystal field
split final state of $f_{5/2}$ state. Distinct signatures of the
Kondo features appearing due to different final states is remarkable
and provides incentive to adopt such procedure to study the
electronic structure of Kondo lattice systems. Ce 4$f$ weight has
been extracted in the past by using different photon energies for
the measurements and exploiting the dependence of photoemission
cross section on photon energy.\cite{patil,Ehm} Since the electronic
states in the valence band are strongly hybridized, they often
exhibit exotic behavior due to their uniqueness different from the
pure elemental spin-orbitals. Therefore, finding out correct
cross-section of these states is difficult. In the present method,
such difficulties can be avoided, although the subtle change in the
structural parameters in different compositions can modify the
uncorrelated electronic structure. Such changes are very small in
the present case and can be neglected within the first
approximation.

\begin{figure}
 \vspace{-2ex}
\includegraphics [scale=0.4]{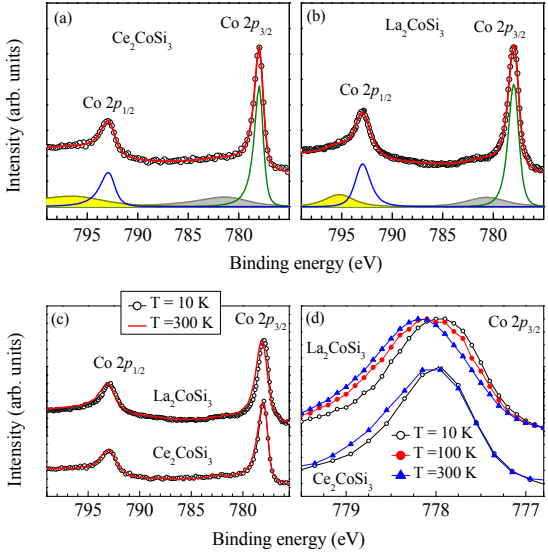}
\vspace{-18ex}
 \caption{Co 2$p$ core level spectra for (a) Ce$_2$CoSi$_3$ and (b)
La$_2$CoSi$_3$. The experimental spectra (symbols) are fitted with
asymmetric line shapes (lines). The thick solid line superimposed on
the symbols are the total fit data. (c) Co 2$p$ spectra of
Ce$_2$CoSi$_3$ and La$_2$CoSi$_3$ at 10 K (open circles) and 300 K
(lines). (d) Expanded view of the Co 2$p_{3/2}$ signal at 10 K (open
circles), 100 K (closed circles) and 300 K (triangles).}
 \vspace{-2ex}
\end{figure}

Since, the photoemission spectroscopy reflects the spectral function
corresponding to the final states of the photo-excitation process,
the presence and absence of the Kondo feature in the electronic
structure can also me manifested in the core level
spectra.\cite{patil} This has been investigated in Fig. 5, where we
show the Co 2$p$ core level spectra of Ce$_2$CoSi$_3$ in Fig. 5(a)
and La$_2$CoSi$_3$ in Fig. 5(b). The experimental spectra exhibit
large asymmetry towards higher binding
energies.\cite{Doniach-Sunjic} While such asymmetry can be
attributed to the low energy excitations across the Fermi level in a
metallic system, the observed spectral lineshape cannot be captured
by simple asymmetric spectral function as shown in the figure. A
broad additional feature needs to be considered in both the cases.
The presence of this feature is distinct in the case of
La$_2$CoSi$_3$ and suggest again the correlated nature of the Co
3$d$ electrons.\cite{fujimoriRMP,sudhir,Schneider}

The decrease in temperature has significant influence in the
spectral functions in both the compounds as shown in Fig. 5(c) and
5(d). While the satellite feature appears to be similar at both 10 K
and 300 K, the main peak exhibits distinct evolutions. In
Ce$_2$CoSi$_3$, the asymmetry of the spectral lineshape is
significantly smaller at 10 K than that at 300 K (see Fig. 5(d)).
This has been attributed to the formation of Kondo singlets at lower
temperatures that leads to larger degree of core hole screening due
to the additional $f$-derived narrow bands near the Fermi level
generated by the Kondo coupling.\cite{patil} Co 2$p_{3/2}$ spectra
of La$_2$CoSi$_3$ exhibit significantly different temperature
evolution - the peak position shifts towards lower binding energies.
The lineshape asymmetry seems to increase gradually with the
decrease in temperature - the La$_2$CoSi$_3$ spectra in Fig. 5(d)
exhibit a sharper rise at low temperatures in the low binding energy
side with gradual fall in intensity at the other side.

\begin{figure}
 \vspace{-2ex}
\includegraphics [scale=0.4]{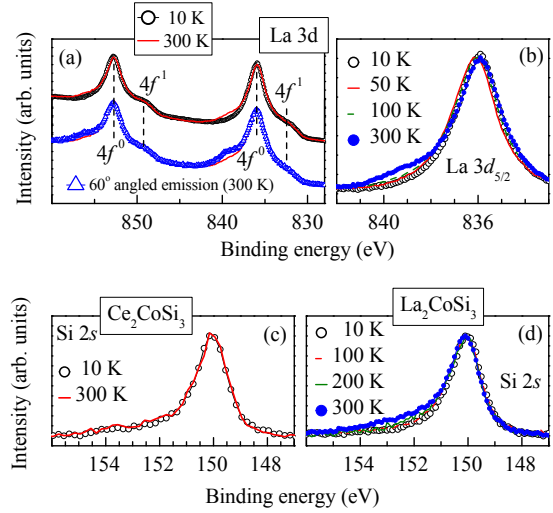}
\vspace{-28ex}
 \caption{(a) La 3$d$ spectra La$_2$CoSi$_3$ collected at 10 K (open
circles) \& 300 K (line) at normal emission, and the 300 K spectrum
collected at 60$^o$ angled emission (open triangles). (b) La
3$d_{5/2}$ photoemission spectra at 10 K (open circles), 50 K (solid
line), 100 K (dashed line) and 300 K (closed circles). (c) Si 2$s$
spectra of Ce$_2$CoSi$_3$ at 10 K (open circles) and 300 K (solid
line). (d) Si 2$s$ spectra of La$_2$CoSi$_3$ at  10 K (open
circles), 50 K (solid line), 100 K (dashed line) and 300 K (closed
circles).}
 \vspace{-2ex}
\end{figure}

The temperature dependence of La 3$d$ core level spectra is plotted
in Fig. 6(a) and 6(b). There are two distinct features for each of
the spin-orbit split lines - the features marked as 4$f^0$ \& 4$f^1$
appearing around 836 eV \& 832.5 eV binding energies for the
3$d_{5/2}$ signal, and 852.8 eV \& 849.3 eV for the 3$d_{3/2}$
signal, respectively correspond to the poorly screened and well
screened final states of the 3$d$ photoemission. We observe an
additional feature about 3.5 eV binding energy higher than the
4$f^0$ peak appearing in both 3$d_{5/2}$ and 3$d_{3/2}$ signals. In
order to identify the origin of this feature, we carried out the
measurements of 3$d$ spectrum at 60$^o$ emission angle, which makes
the technique significantly surface sensitive.\cite{manju} Clearly,
the higher binding energy feature enhances in intensity with the
increase in surface sensitivity indicating its surface character.
Decrease in temperature from 300 K to 10 K does not have significant
influence in the spectral lineshape corresponding to the bulk
electronic structure, while the surface feature appear to decrease
gradually with the decrease in temperature (see Fig. 6(b)) and
becomes almost non-existent at 10 K. This is a unique behavior of
this material exhibiting surface bulk differences in the electronic
structure as the temperature is raised, which can have significant
implication in device applications.

We investigate the Si 2$s$ core level spectra in Fig. 6(c) and 6(d),
for Ce$_2$CoSi$_3$ and La$_2$CoSi$_3$, respectively. Each of the
spectra exhibit sharp single peak structure with large asymmetry
towards higher binding energies. The temperature dependence of Si
2$s$ core level of Ce$_2$CoSi$_3$ shows almost identical lineshape
at 10 K and 300 K. A small increase in the Doniach-\v{S}unji\'{c}
asymmetry at higher binding energies at higher temperatures is
observed in the case of La$_2$CoSi$_3$. The change was inferred to
be of bulk origin through varying the emission angle of the
photoelectrons (not shown here) by 60$^o$ as was done above for the
La 3$d$ core level. The energy shift of the peak position in Si 2$s$
and La 3$d$ spectra is negligible.

Evidently, the core level spectra of La$_2$CoSi$_3$ shows
interesting evolution of the lineshape and peak positions with the
change in temperature. The decrease in asymmetry with the decrease
in temperature can be attributed to the Kondo-resonance behavior due
to Fermi surface reconstructions induced by the Kondo singlets at
low temperatures. While such an effect is not observed in non-Kondo
system, La$_2$CoSi$_3$, this system shows significant anomalies in
all the core levels. Co 2$p$ peak shifts and become more asymmetric
at lower temperatures. A surface feature disappears in the La 3$d$
spectra and the asymmetry in Si 2$s$ reduces at low temperatures.
All these effects, may be related to the emergence of surface-bulk
differences in the electronic structure at higher temperatures,
reveals complexity of the problem.

\section{conclusions}

In summary, we studied the electronic structure of La$_2$CoSi$_3$
using high resolution photoemission spectroscopy and compared it
with the band structure calculations. The valence band spectra of
La$_2$CoSi$_3$ showed Fermi liquid evolution of the spectral
intensity close to $\epsilon_F$. The symmetrized He {\scriptsize II}
spectra does not show a change in the spectral density of states as
a function of temperature in this system. The comparison of the
valence band spectra with the simulated ones from the LDA + $U$
results suggests the existence of electron correlations among the Co
3$d$ electrons and the corresponding Hubbard, $U$ is close to 3 eV.
The Co 2$p$ core level spectra exhibit significantly intense
satellite features providing further evidence of the correlation
induced effects in the electronic structure. We show that the
spectral functions of this material can be utilized to extract the
Ce 4$f$ contributions from the Ce$_2$CoSi$_3$ spectra. Comparison of
the temperature evolution of Co 2$p$ lineshape is found to be
different in Kondo and non-Kondo systems, which may be utilized as a
signature of Kondo behavior. The La 3$d$ core spectra reveal a
surface feature as the temperature is raised suggesting emergence of
surface-bulk differences at finite temperatures. The Co 2$p$ and Si
2$s$ peaks exhibit curious evolution with temperature. The anomalies
in the core level spectra may have relation to the emergence of the
surface-bulk differences at higher temperatures.

\section{Acknowledgements}

The author, S. P. thanks the Council of Scientific and Industrial
Research, Government of India for financial support. All the authors
thank Mr. Kartik K. Iyer for his help in sample preparation and
characterization.

\end{document}